\documentclass[%
 reprint,
 amsmath,amssymb,
prb,
]{revtex4-2}

\usepackage{graphicx}
\usepackage{dcolumn}
\usepackage{bm}
\usepackage{hyperref}
\usepackage{cleveref}
\usepackage{physics}
\usepackage{siunitx}
\usepackage{todonotes}

\crefformat{footnote}{#2\footnotemark[#1]#3}

\begin{document}

\preprint{APS/123-QED}

\title{Dark Exciton Giant Rabi Oscillations with no External Magnetic Field}

\author{Vladimir Vargas-Calderón}
\email{vvargasc@unal.edu.co}
\author{Herbert Vinck-Posada}%
\affiliation{%
Grupo de Superconductividad y Nanotecnología, Departamento de Física\\
Universidad Nacional de Colombia, 111321, Bogotá, Colombia
}%

\author{J. M. Villas-Boas}
\affiliation{
Instituto de Física, Universidade Federal de Uberlândia, 38400-902, MG, Brazil
}%

\date{\today}

\begin{abstract}
Multi-phonon physics is an emerging field that serves as a test bed for fundamental quantum physics and several applications in metrology, on-chip communication, among others. Quantum acoustic cavities or resonators are devices that are being used to study multi-phonon phenomena both theoretically and experimentally. In particular, we study a system consisting of a semiconductor quantum dot pumped by a driving laser, and coupled to an acoustic cavity. This kind of systems has proven to yield interesting multi-phonon phenomena, but the description of the quantum dot has been limited to a two-level system. This limitation restrains the complexity that a true semiconductor quantum dot can offer. Instead, in this work we consider a model where the quantum dot can have both bright and dark excitons, the latter being particularly useful due to their lower decoherence rates, because they do not present spontaneous photon emission. In this setup, we demonstrate that by fine-tuning the driving laser frequency, one is able to realise giant Rabi oscillations between the vacuum state and a dark exciton state with $N$-phonon bundles. From this, we highlight two outstanding features: first, we are able to create dark states excitations in the quantum dot without the usual external magnetic field needed to do so; and second, in a dissipative scenario where the acoustic cavity and the quantum dot suffer from losses, the system acts as a phonon gun able to emit $N$-phonon bundles.
\end{abstract}

\maketitle


\section{Introduction}\label{sec:introduction}
Quantum vibrational modes of solids, described by phonons, have a great potential to be used in technological applications in metrology or quantum information processing~\citep{balandin2005nanophononics,manenti2017circuit,zhang2018noon,schuetz2015,noguchi2017,arrangoiz2018}. Moreover, as the control of individual phonons continues to improve, the study of phonons and their interaction with other excitations in many-body quantum systems is also relevant for testing fundamental physics~\citep{ask2019,vonlupke2021parity,reiter2019,Wei2018}. Analogous to photons in quantum electrodynamics, phonons can be used to store, process, and transduce quantum information. The inclusion of phonons to the quantum toolbox gives a three-fold advantage: first, losses by radiation into the electromagnetic field vacuum are no longer present, as phonons can only propagate through some material medium (usually in solid-state devices); second, phonon energy scales are, in general, different from the optical energy scale, making phonons especially suited for on-chip communication~\citep{schuetz2015,manenti2017,bin2020n,wan2021fault} in a variety of characteristic energies, from MHz to THz~\citep{bell1976surface,weig2004,rozas2009,soykal2011,fainstein2013,kharel2018ultra,bolgar2018,chen2019,gokhale2020,sandeep2018,zalalutdinov2021,gregory2020,yiwen2017,conell2010,santos2018,pirkkalainen2013,lahaye2009,martin2014,ockeloen2017,massel2011,Rouxinol2016,li2008,stock2011,wigger2021,VARTANIAN2018548,kettler2021}; third, many experimental techniques developed by solid-state physicists~\citep{ask2019,xu2018high,kharel2018ultra,akimov2017review} become available for quantum information processing tasks with phonons~\citep{manenti2017,wan2021fault}.

Even though the majority of theoretical and experimental efforts have been devoted to single-phonon generation and control, many-phonon states are also required for highly non-classical sources, useful for quantum sensing and metrology~\citep{zhang2018noon}, and quantum technologies such as quantum memories and transducers~\citep{schuetz2015,noguchi2017,arrangoiz2018}. To this end, the recent work by~\citet{bin2020n} proposes a physical system in quantum acoustodynamics composed of a semiconductor quantum dot (QD), modelled as a two-level system, coupled to the phonon mode of an acoustic cavity. The QD is coherently pumped by an external laser with a frequency that can be tuned to excite giant Rabi oscillations between a state that is mostly the vacuum state (no QD excitations and 0 phonons in the acoustic cavity), and a state that is composed of an exciton QD state, accompanied by $N$ phonons in the acoustic cavity. The authors then show that dissipative channels allow the emission of $N$-phonon bundles, and analyse the quantum statistics of this emission, finding out that depending on dissipative and Hamiltonian parameters, they can realise a phonon laser or a phonon gun.

In this work we study a QD coherently pumped by an external laser and immersed in an acoustic single-mode cavity~\citep{Wei2018}, but we take into account the richer exciton basis provided by the different spin alignments of the electron and hole that compose the exciton. Thus, the QD is described by a ground state, two bright exciton states, and two dark exciton states. The inclusion of this richer exciton basis allows the description of more complex interactions in the system, which enables the control of interesting phenomena. Therefore, the main contributions of our work, coming from the more complete exciton basis, are: dark excitons can be excited, taking advantage of the Bir-Pikus interaction~\citep{bir1974symmetry}, by fine-tuning the laser frequency without the usual need for an external magnetic field~\citep{carlos2017,neumann2021} --which is both experimentally challenging as well as expensive~\citep{carlos2017,adambukulam2021ultra}; and giant-Rabi oscillations between a vacuum state and a dark exciton-$N$-phonon state can be realised to emit $N$-phonon bundles.

We highlight that the laser frequency fine-tuning can be performed to target any desired giant Rabi oscillation for a wide range of parameters such as decay rates, characteristic energies and coupling constants. 
Therefore, the method that we present can be seen as a recipe to produce giant Rabi oscillations. 
The damping of those oscillations, though, will be determined by the dissipative nature of the physical realisation of the embedded QD in an acoustic cavity.

The paper is organised as follows.
In~\cref{sec:model} we introduce the Hamiltonian that models the physical system under consideration.
The analysis and discussion of $N$-phonon bundle emission is presented in~\cref{sec:results}.
We draw conclusions in~\cref{sec:conclusions}.
Throughout the paper, we use parameters similar in magnitude to those presented in Ref.~\citep{bin2020n}, as this allows us to build a bridge between our results and theirs, allowing a direct comparison of the phenomenology that we find.
Nonetheless, some of these parameters are orders of magnitude away from the state of the art experiments involving the coupling of QDs to acoustic cavities. 
This does not invalidate the rich physics that is described.
In fact, as we pointed out, the recipe to produce giant Rabi oscillations is found to hold for more experimentally feasible parameters, as we exemplify and discuss in~\cref{sec:feasibility}.


\section{Model}\label{sec:model}
\begin{figure}
    \centering
    \includegraphics[width=\columnwidth]{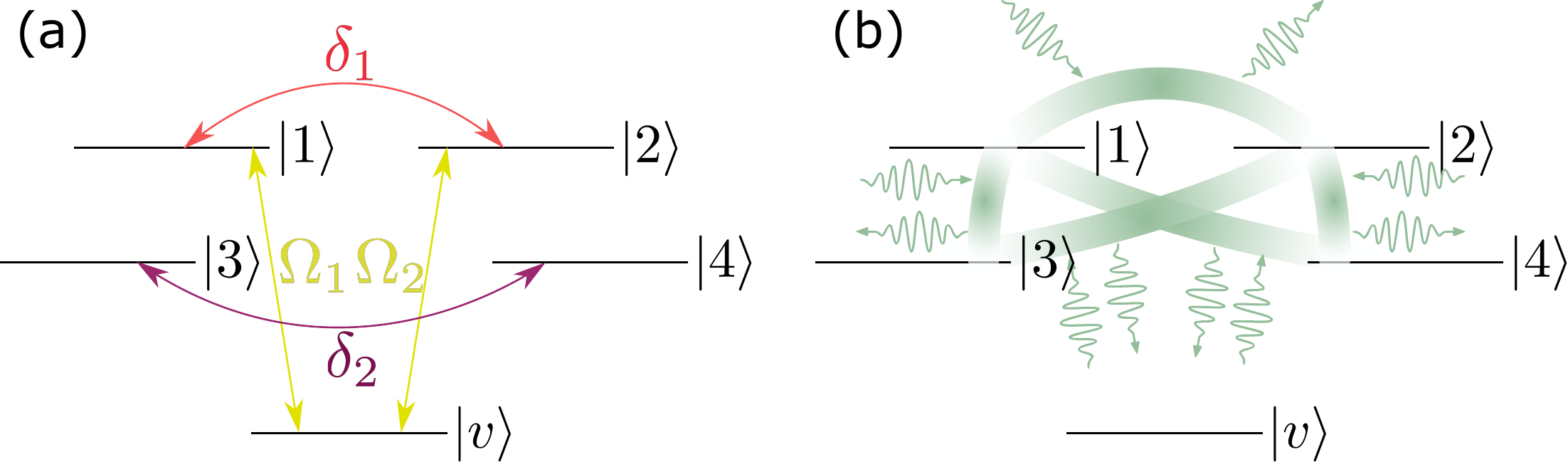}
    \caption{Interactions that cause transitions between different QD states. (a) shows the interactions portrayed in $H_\text{QD}$ and $H_\text{laser}$ with two-directional arrows, and (b) shows the interactions portrayed by $H_\text{el-ph}$ with gradient-coloured lines, which are mediated by phonon absorption and emission processes in the acoustic cavity.}
    \label{fig:qdlevels}
\end{figure}

In this work we consider a pumped quantum dot embedded in an acoustic cavity described by the following Hamiltonian~\citep{bin2020n}:
\begin{align}
    H = H_\text{QD} + H_\text{laser} + H_\text{cav} + H_\text{el-ph}.\label{eq:wholehamiltonian}
\end{align}
The bare quantum dot is described by a valence state $\ket{v}$, two bright exciton states $\ket{1}$ and $\ket{2}$ with anti-parallel electron-hole spins, and two dark exciton states $\ket{3}$ and $\ket{4}$ with parallel electron-hole spins. The corresponding Hamiltonian, taking into account the exchange interaction~\citep{nomura1994,bayer2002}, and making use of the ladder operators $\sigma_{ij} = \ketbra{i}{j}$, reads~\citep{carlos2017}
\begin{align}
\begin{aligned}
    H_\text{QD} ={}& \omega_X (\sigma_{11} + \sigma_{22}) + \omega_d(\sigma_{33} + \sigma_{44})\\ &
    + \frac{\delta_1}{2}(\sigma_{12} + \sigma_{21}) + \frac{\delta_2}{2}(\sigma_{34} + \sigma_{43})
\end{aligned},\label{eq:qdhamiltonian}
\end{align}
where $\omega_X$ is the bare bright exciton energy, $\omega_d=\omega_X - \delta_0$ is the shifted bare dark exciton energy, and $\delta_{1(2)}$ split the bright(dark) exciton energies. Further, the QD is driven by an external laser that pumps the bright exciton states through $H_\text{laser} = \Omega_1 (e^{-i\omega_L t}\sigma_{1v} + e^{i\omega_L t}\sigma_{v1}) + \Omega_2(e^{-i\omega_L t}\sigma_{2v} + e^{i\omega_L t}\sigma_{v2})$. Here, the relative magnitudes of the laser amplitudes $\Omega_1$ and $\Omega_2$ depend on the laser polarisation~\citep{belhadj2009}. The time dependence of the whole Hamiltonian is removed in the laser-frequency rotating frame via the unitary transformation $U=\exp(i\omega_Lt[\sigma_{11}+\sigma_{22}+\sigma_{33}+\sigma_{44}])$. The acoustic cavity Hamiltonian accounts for the single phonon mode energy $H_\text{cav} = \omega_b b^\dagger b$, where $b$ is the phonon annihilation operator. Throughout the paper we will use the values $\delta_0=0.04\omega_b$, $\delta_1=0.036\omega_b$ and $\delta_2=0.01\omega_b$, which match the values reported by~\citet{bayer2002} for $\omega_b=5\si{\milli\electronvolt}$. Finally, we also consider the coupling of the hole spin to the strain tensor of the QD--described by the Bir-Pikus Hamiltonian~\citep{bir1974symmetry,woods2004}--as well as the electron-hole exchange interaction~\citep{takagahara1993}, resulting in an electron-phonon Hamiltonian that is able to generate phonon-mediated transitions between bright and dark excitons, and between the two bright excitons~\citep{roszak2007}
\begin{align}
\begin{aligned}
    H_\text{el-ph} = \left\{\frac{g_\text{bd}}{\sqrt{2}}\left[(1+i)(\sigma_{13} + \sigma_{14}) + (1-i)(\sigma_{23} + \sigma_{24})\right]\right.\\
    \left.\phantom{\frac{a}{2}}+g_\text{bb}[\sigma_{11} + \sigma_{22} + i(\sigma_{12} - \sigma_{21})]\right\}(b^\dagger + b) + \text{H.c.}
\end{aligned}
\end{align}
where $g_\text{bb(bd)}$ are bright-bright(bright-dark) exciton coupling rates through phonons. An illustration of the interactions that affect the QD state is given in~\cref{fig:qdlevels}.

\section{Results}\label{sec:results}
\begin{figure}[ht]
    \centering
    \includegraphics[width=\columnwidth]{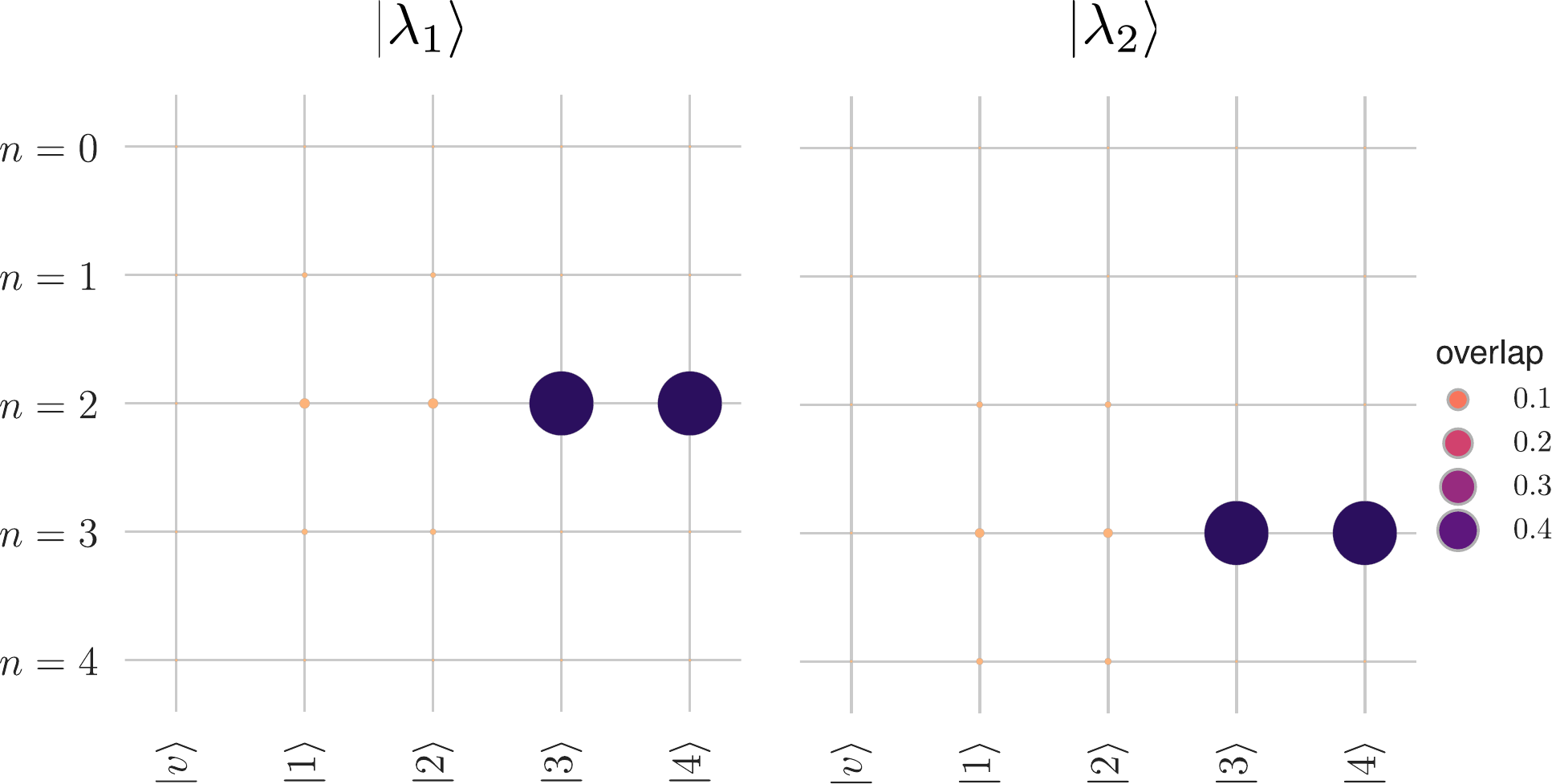}
    \caption{Eigenstates of~\cref{eq:wholehamiltonian} for weak electron-phonon coupling ($g=g_\text{bb}=g_{\text{bd}} \approx 0.02  \omega_b$), pumping ($\Omega_1\approx0.082 \omega_b$, $\Omega_2=0$) and the external laser in resonance with the bright exciton energy ($\omega_L = \omega_X$, where the frequency of the QD can be chosen to match any semiconductor QD exciton energy, e.g. $1.36\si{\electronvolt}$ for GaAs QDs). These values are used throughout the paper, except for the laser frequency, which will be fine-tuned. The state $\ket*{\lambda_{1(2)}}$ is mostly composed by the state $\ket{2(3),d_+}$, where $\ket{d_\pm} = (\ket{3} \pm \ket{4})/\sqrt{2}$ is the dark exciton symmetric ($+$) or anti-symmetric ($-$) state.}
    \label{fig:darkeigenstates}
\end{figure}

Such an electron-phonon coupling, accompanied by the optically driven Stokes process through the laser pumping, allows the excitation of giant Rabi oscillations between the vacuum state $\ket{\text{phonons}=0}\otimes\ket{\text{QD}=v}$ (for weak laser pumping) and an eigenstate of the Hamiltonian shown in~\cref{eq:wholehamiltonian}~\citep{bin2020n}. In particular, we can select an eigenstate mainly composed of $N$-phonons and dark excitons. As an example, we show in~\cref{fig:darkeigenstates} two such eigenstates that will be used as targets for giant Rabi oscillations.

\begin{figure}[ht]
    \centering
    \includegraphics[width=\columnwidth]{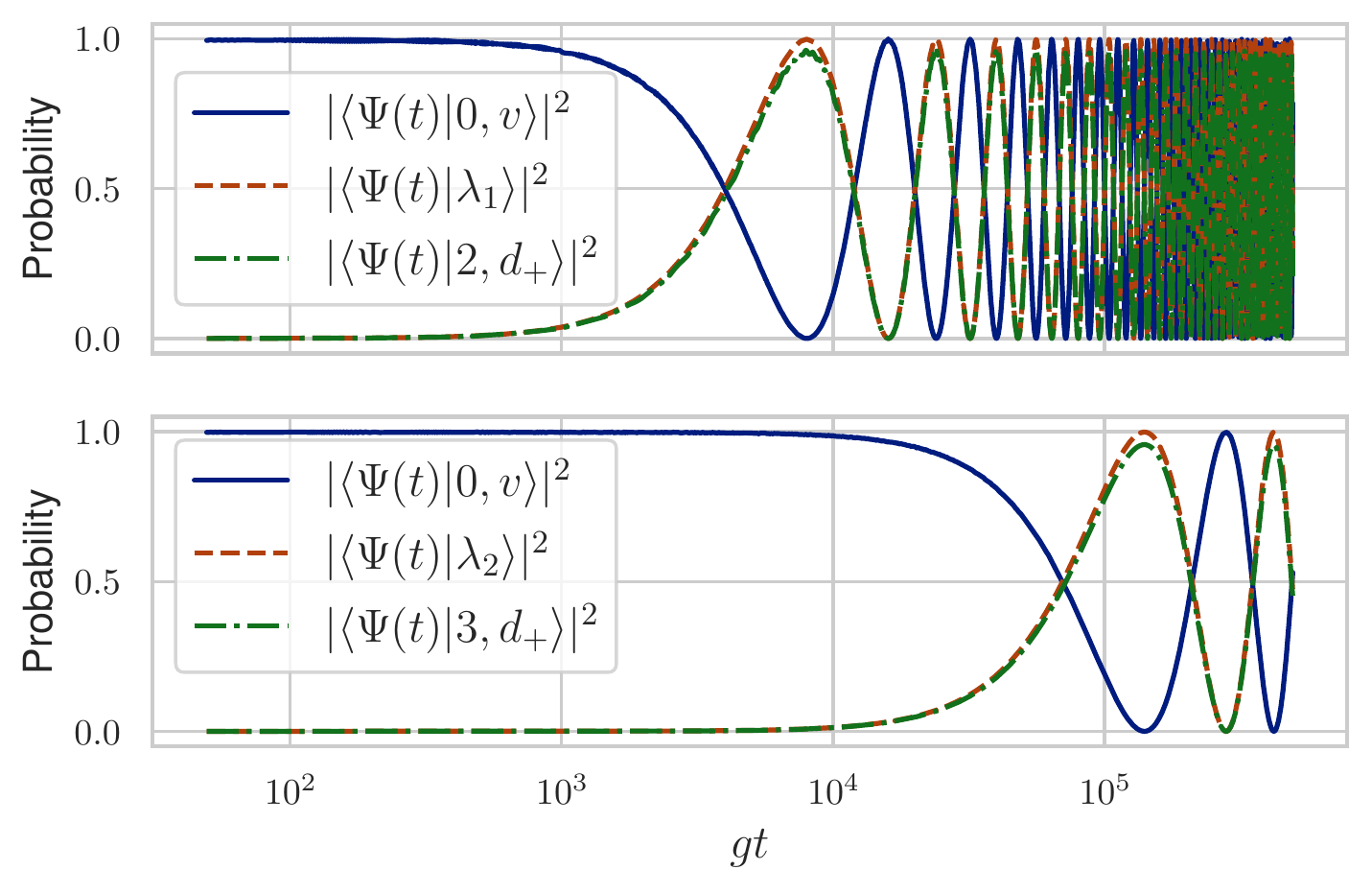}
    \caption{Giant Rabi oscillations between the 0-phonon valence state (solid line) and the states shown in~\cref{fig:darkeigenstates} (dashed line). The dash-dotted line shows the evolution of the $\ket{n, d_+}$ state for $n=2$ (top panel) and $n=3$ (bottom panel). The laser detuning is $\Delta/\omega_b = (\omega_X - \omega_L)/\omega_b\approx-1.960$ for the top panel, and ${\approx}-2.961$ for the bottom panel. The same coupling and pumping conditions from~\cref{fig:darkeigenstates} are used.}
    \label{fig:superRabi}
\end{figure}

Giant Rabi oscillations are achieved through a cascading effect~\citep{bin2020n}, where the system transitions from the vacuum state $\ket{0,v}$ to a QD bright exciton state (depending on the laser amplitudes), and then to a (mostly) dark exciton symmetric state with $N$ phonons after the system is guided by the electron-phonon coupling mechanism. Examples of such giant Rabi oscillations involving the 2- and 3-phonon states shown in~\cref{fig:darkeigenstates} can be found in~\cref{fig:superRabi}. Both examples show that it is possible to tune the laser frequency to target giant Rabi oscillations between the vacuum state $\ket{0,v}$, and any eigenstate $\ket{\lambda}$ of~\cref{eq:wholehamiltonian} by considering a laser frequency $\omega_L = \omega_\lambda - \omega_g$, where $\omega_\lambda = \bra{\lambda}H\ket{\lambda}$ and $\omega_g$ is its ground state frequency~\footnote{Note that the ground state is not exactly $\ket{0,v}$, because the states are dressed by the external laser.}. It is worth noting that the states $\ket{n,d_-}$, i.e. with anti-symmetric dark exciton matter states, are eigenstates of~\cref{eq:wholehamiltonian}, but they are not accessible from non-dark anti-symmetric initial states, as can be easily corroborated by computing $\bra{m, \beta} H \ket{n, d_-}=0$, where $\ket{\beta}$ is any state spanned by bright exciton states, and $m$ is some number of phonons. However, in the presence of dissipation they can be accessed, as will be explored later.

The Hamiltonian analysis so far presented is only relevant to understand the underlying physics of giant Rabi oscillation production, but to study a realistic setup of the proposed system it is necessary to describe it as an open quantum system. Such a description shows that giant Rabi oscillations can emit $N$-phonon bundles through dissipative channels. In the case of weak coupling to the environment, which leads to the usual Born-Markov approximations~\citep{breuer2002theory}, we consider four dissipative channels under the Gorini–Kossakowski–Sudarshan–Lindblad equation for the system's density operator~\citep{lindblad1976generators,gorini1976completely}
\begin{align}
\begin{aligned}
    \frac{d\rho}{dt} ={}& i[\rho, H] + \kappa \mathcal{D}_b[\rho] + \gamma_b\sum_{j=1,2}\mathcal{D}_{\sigma_{vj}}[\rho] \\&+ \gamma_d\sum_{j=3,4}\mathcal{D}_{\sigma_{vj}}[\rho] + \gamma_\phi\sum_{j=1}^4\mathcal{D}_{\sigma_{jj}}[\rho]
\end{aligned},\label{eq:lindblad}
\end{align}
where $\mathcal{D}_A[\rho] = A\rho A^\dagger - \frac{1}{2}\rho A^\dagger A - \frac{1}{2} A^\dagger A \rho$ is the dissipator super-operator of the collapse operator $A$. The four dissipative channels considered in~\cref{eq:lindblad} are: phonon escape from the acoustic nanocavity at a rate $\kappa$ with a collapse operator $b$ due to unwanted coupling with leaky modes~\citep{perea2004dynamics,winter2007}, spontaneous emission of the bright excitons at a rate $\gamma_b$ with collapse operators $\sigma_{vj}$ for $j=1,2$, effective spontaneous emission of the dark excitons (hole or electron spin flip followed by bright exciton spontaneous emission~\citep{crooker2003multiple}) at a rate $\gamma_d$ with collapse operators $\sigma_{vj}$ for $j=3,4$, and pure dephasing of all exciton states at a rate $\gamma_\phi$ with collapse operators $\sigma_{jj}$~\citep{takagahara2002theory}.

\begin{figure}[ht]
    \centering
    \includegraphics[width=\columnwidth]{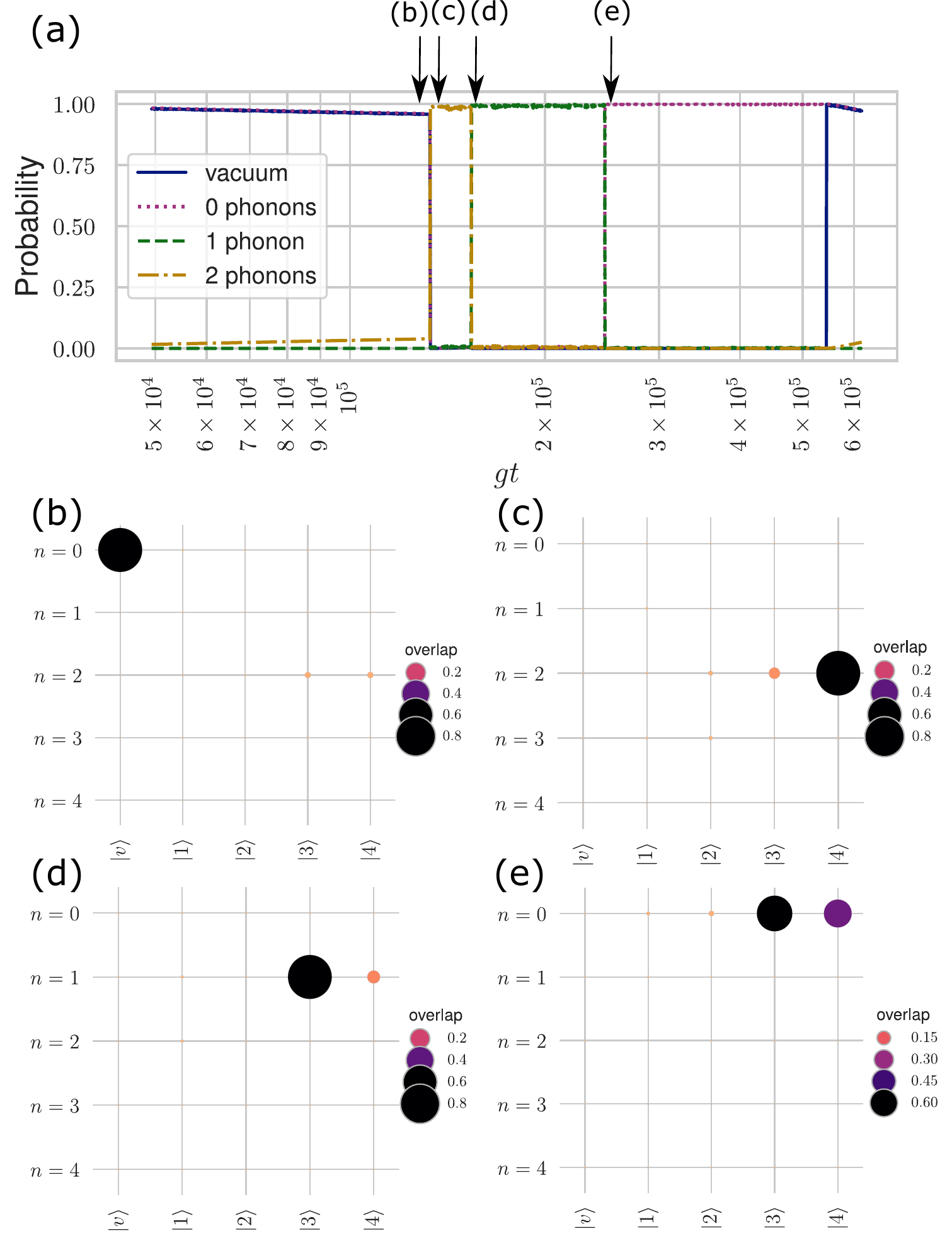}
    \caption{Evolution of a quantum trajectory due to Hamiltonian dynamics and dissipation-induced quantum jumps. (a) shows the occupation of the vacuum $\ket{0,v}$ state, and of the $N$-phonon states, i.e. the occupation of $\sum_i\ket{N,i}$. (b)-(d) show the state composition of the system at the times pointed out by an arrow in panel (a). The laser detuning is the same as in the top panel of~\cref{fig:superRabi}. The dissipative parameters are $\kappa \approx 7.89\times10^{-4}\omega_b$, $\gamma_b \approx 1.869\times 10^{-5}\omega_b$, $\gamma_d = 0.1\gamma_b$ and $\gamma_\phi = 4\times 10^{-4}\omega_b$. These parameters are used throughout the paper.}
    \label{fig:emission}
\end{figure}

Solving~\cref{eq:lindblad} using quantum trajectories~\citep{dalibard1992,molmer1993monte} exposes the $N$-phonon bundle nature of the excitations that are emitted: the $N$-phonon bundle behaves as a quasi-particle in the context of the dynamical process of emission~\citep{munoz2014emitters}.
\Cref{fig:emission} depicts this concept by showing a single quantum trajectory that suffers a strongly-correlated phonon emission process. 
The initial state is the vacuum, as shown in~\cref{fig:emission}(b), right before a quantum jump that takes the system to a dark state with two phonons, as shown in~\cref{fig:emission}(c).
This jump is due to the pure dephasing dissipative channel with a quantum jump operator $\sigma_{44}$. The state in~\cref{fig:emission}(b) is, to a good approximation, $c_+(t)\ket{0,v} + c_-(t)\ket{2,d_+}$, where $\abs{c_+(t)}^2 + \abs{c_-(t)}^2 = 1$.
Before the quantum jump, $\abs{c_+(t)}^2 \approx 1$, meaning that the system is still in the vacuum state.
The dephasing jump takes the system into a state of two phonons and a dark state, as shown in~\cref{fig:emission}(c).
Now, the phonon dissipative channel--with jump operator $b$--makes the system undergo a quantum jump through the emission of a phonon, leaving the system in a dark state, but with only one phonon, as shown in~\cref{fig:emission}(d).
The quantum trajectory simulation shows that yet another quantum jump occurs between the points (d) and (e) due to dephasing. 
This jump, however, does not affect the phonon population.
Although the phonon-escape mechanism depletes the cavity phonon-by-phonon, it is shortly after the emission of the first phonon that the second phonon is also emitted due to the jump operator $b$, leaving the system in a dark excitonic state with zero phonons, as shown in~\cref{fig:emission}(e).
After the point (e), there are a couple of further dephasing quantum jumps that leave the phonon population unaffected because the exciton already occupies the dark states.
Finally, the QD emits a photon due to recombination (jump operator $\sigma_{v3}$, which models a random spin flip followed by spontaneous emission), and the system is allowed to transit the same process once again. 
The two-phonon bundle emission process arises in all quantum trajectories, and becomes more common when the electron-phonon coupling $g$ is increased, at the cost of realising giant Rabi oscillations between the vacuum and states that have a larger bright contribution. 
Nonetheless, we point out that for the presented parameters, one-phonon emission processes are more common~\footnote{Parameters can be tuned to get a larger amount of two-phonon emission processes. In this work, we tuned parameters with the tree-structured Parzen estimator implemented in Optuna~\citep{optuna2019} to minimise the generalised second-order two-phonon bundle correlation function at a value for the laser detuning that coupled the vacuum and the $\ket{2,d_+}$ state through giant Rabi oscillations.}.

\begin{figure}[ht]
    \centering
    \includegraphics[width=\columnwidth]{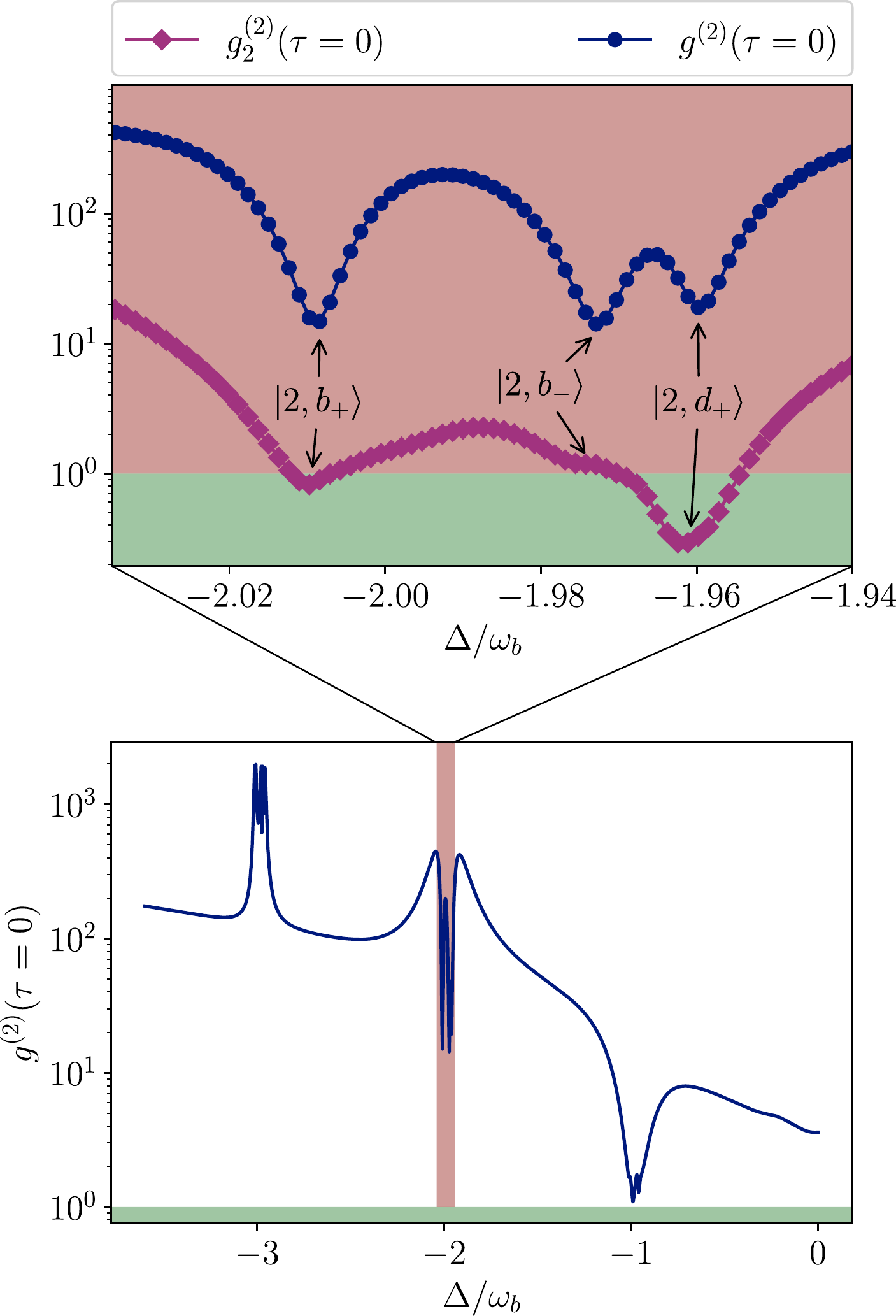}
    \caption{Equal-time one- and two-phonon bundle correlation functions. At $\Delta\approx -2\omega_b$ there are dips in the super-bunching peak of the usual correlation function at resonances corresponding to two-phonon states with bright (anti-)symmetric $\ket{b_{(-)+}}$, and with the dark symmetric state $\ket{d_+}$. The red/green-shaded regions align with sub-/super-Poissonian statistics regions.}
    \label{fig:g2}
\end{figure}

Moreover, as in the work by \citet{bin2020n}, a signature of strong correlations of the emitted phonons is found in the equal-time $n$-th order phonon correlation function $g^{(n)}(\tau=0) = \expval*{b^{\dagger n}b^n}/\expval*{b^\dagger b}^n$~\citep{dell2006multiphoton}.
They noticed that near the resonances $\Delta \approx -n\omega_b$ (for $n\ge 2$), this correlation function did not show the super-bunching peaks that would intuitively be expected for multi-phonon emission.
Instead, there are dips in the middle of such peaks, as we show in~\cref{fig:g2}. In their case, only one dip coinciding with the resonance with the $\ket{n,c}$ state was shown because their model was restricted only to one conduction $\ket{c}$ exciton state, whereas in our case, we present three different dips corresponding to the resonances with the bright symmetric and anti-symmetric states, as well as the dark symmetric state.

However, such correlation function is not adequate to investigate multi-phonon properties associated to the emitted $N$-phonon bundles.
This is due to the fact that the $g^{(2)}(\tau=0)$ correlation function--normally measured in photonic setups such as the Hanbury Brown and Twiss interferometer~\citep{brown1956}--is related to the co-occurrence of single phonon detection events, as originally derived by~\citet{glauber1963}.
On the other hand, the equal-time $m$-th order $N$-phonon bundle correlation function, defined as $g^{(m)}_N(\tau=0) = \expval*{b^{\dagger Nm} b^{Nm}}/\expval{b^{\dagger N}b^N}^m$ correctly describes these multi-phonon properties because it treats the $N$-phonon bundle as a quasi-particle~\citep{munoz2014emitters}, with associated creation and annihilation operators $b^{\dagger N}$ and $b^N$, respectively.
In~\cref{fig:g2} we show that this generalised correlation function reaches the sub-Poissonian anti-bunching regime for some values of the laser detuning~\citep{zou1990}.
The resonance associated to the dark symmetric with two phonons state presents the lowest value of $g_2^{(2)}$, well within the anti-bunching regime.
This shows that the Bir-Pikus mechanism allows the generation of robust anti-bunching behaviour when the laser frequency is tuned to target giant Rabi oscillations with the dark symmetric state, even more so than with the usually accessed bright states.

\begin{figure}[ht]
    \centering
    \includegraphics[width=\columnwidth]{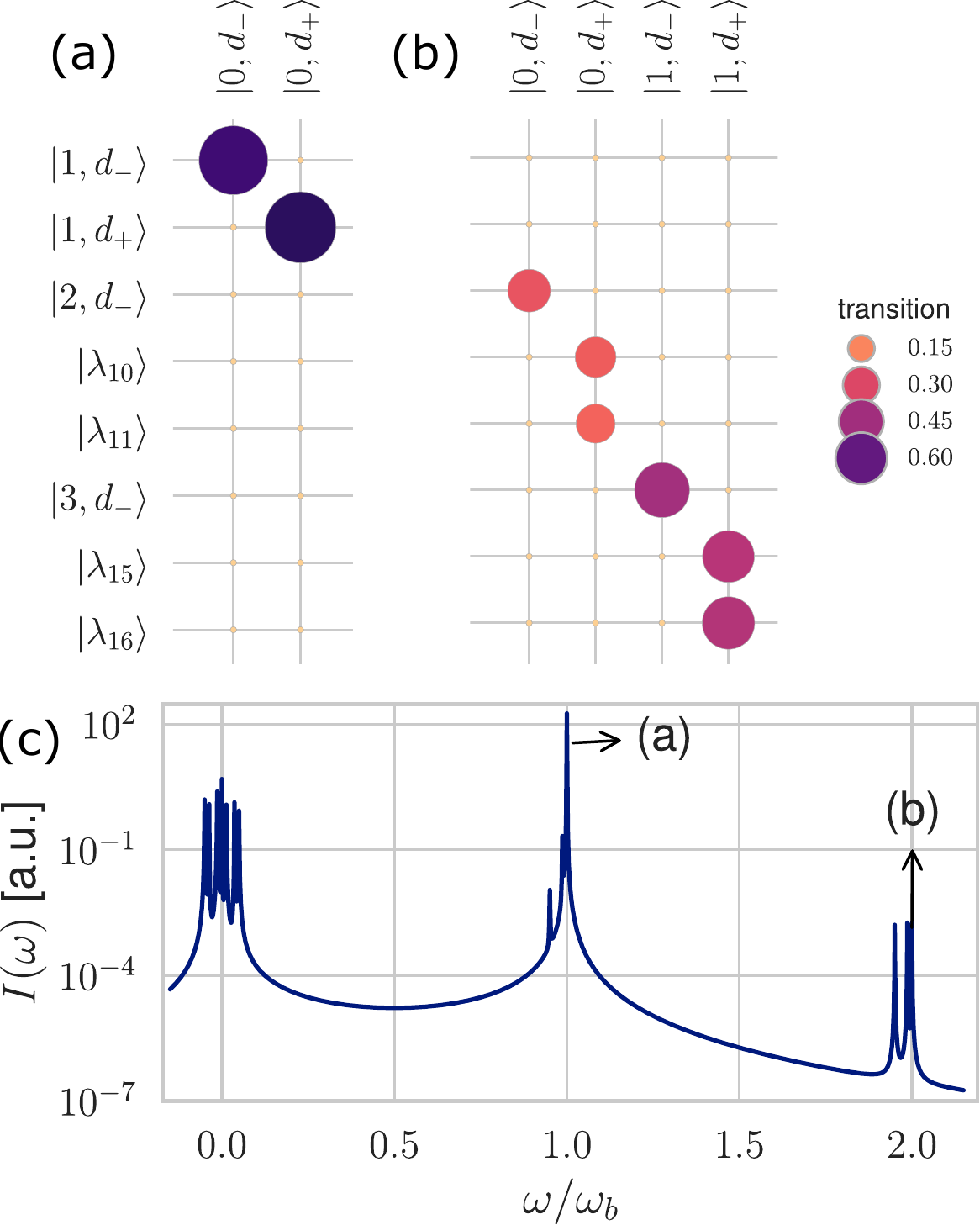}
    \caption{Phonon emission spectrum $I(\omega)$. Panels (a) and (b) show the elements $\abs{\bra{\psi}\varrho \ket{\phi}}^2$ of the Liouvillian eigenmatrices $\varrho$ that match the peaks pointed in the phonon emission spectrum in panel (c). The eigenmatrices show that the peak (a) corresponds to a one phonon emission process, whereas the peak (b) corresponds to two-phonon emission processes. Eigenstates $\ket{\lambda_{10}}$ and $\ket{\lambda_{11}}$ of \cref{eq:wholehamiltonian} are superpositions of $\ket{0,v}$ and $\ket{2,d_+}$, and eigenstates $\ket{\lambda_{15}}$ and $\ket{\lambda_{16}}$ are superpositions of $\ket{0,v}$ and $\ket{3,d_+}$. The laser detuning is the same as in the top panel of~\cref{fig:superRabi}.}
    \label{fig:spectrum}
\end{figure}

Finally, we examine the phonon emission spectrum (see~\cref{fig:spectrum}(c)) to show that the two-phonon emission processes can be frequency-resolved.
The phonon emission spectrum $I(\omega)$ can be obtained through the Wiener-Khintchine theorem in analogy to the photoluminiscence spectrum: $I(\omega) \propto \frac{\kappa}{\pi}\int_0^\infty \expval{b(t)b^\dagger(t+\tau)}e^{i\omega\tau}\,d\tau$~\citep{mollow1969,perea2004dynamics,VARGASCALDERON2019168}.
We can interpret every single peak through the spectral theory of the Liouvillian superoperator $\mathcal{L}$ that satisfies~\cref{eq:lindblad}, written as $\frac{d\rho}{dt} = \mathcal{L}[\rho]$~\citep{petrosky1996,petrosky2010,manzano2018}.
A formal solution of this equation for a time-independent Liouvillian is $\rho(t) = \sum_k e^{\Lambda_k t} \Tr[\varrho_k \rho(0)]\varrho_k$, where $\Lambda_k$ are the complex eigenvalues of $\mathcal{L}$ with corresponding eigenmatrices $\varrho_k$. 
The eigenvalues are associated to the emission peaks~\citep{albert2014,naomichi2015,vargasphonon} as they show both the peak location $\Im{\Lambda_k}$ and the full width at half maximum $-\Re{\Lambda_k}$. 
On the other hand, the eigenmatrices account for information about which states are involved in each transition~\citep{tay2008biorthonormal}. 
These transitions can be systematically studied when they involve changes in the number of excitations through dissipative processes only~\citep{torres2014,echeverri2018,VARGASCALDERON2020126076}. 
In our case, such a theory cannot be used because of the presence of coherent pumping to the QD. 
Nonetheless, the examination of the eigenmatrices matrix elements still shows which states are involved in the transitions, as displayed in~\cref{fig:spectrum}(a) and (b).

\Cref{fig:spectrum}(a) shows that the most prominent peak of the spectrum matches transitions between dark states with one phonon and dark states with 0 phonons. It is worth noting that the magnitude of the matrix elements shown in~\cref{fig:spectrum}(a) and (b) do not indicate the contribution of those transitions to the emission peak. Instead, the excitation of the transitions depends upon the energy injected into the system, and the strength of the interactions that enable certain population transfers. Further, \cref{fig:spectrum}(b) shows the allowed transitions for another peak, at a frequency $\omega=2\omega_b$, which matches transitions between dark states with two phonons decaying to dark states with no phonons, and also dark states with three phonons decaying to dark states with one phonon. We highlight that the emission peak coinciding with the two-phonon emission is much smaller than the one-phonon emission peak. However, the corresponding transitions can be differentiated due to the large energy difference in the spectrum. For instance, an analogous procedure as the one presented by~\citet{SanchezMunoz2018} can be proposed for phonons. The associated side-bands on each major peak are due to transitions between dark and bright states accompanied by zero-, one- and two-phonon emission.

\section{Conclusions}\label{sec:conclusions}
Dark excitons in QDs are more robust to decoherence than bright excitons, because they cannot couple to leaky optical modes. It is usually assumed that dark excitons are produced (and controlled) via external magnetic fields. However, in this work, we show that dark excitons in a QD coherently pumped by a laser and coupled with an acoustic cavity not only can be produced, but can also be targeted to realise giant Rabi oscillations between dark states with $N$ phonons and the vacuum. These giant Rabi oscillations show a restricted cascading process that couples states with different phonon numbers, ultimately resulting in an effective coupling between the vacuum and a dark symmetric exciton state with $N$-phonons.

In the driven-dissipative scenario, other intermediate states are activated in the cascading process, culminating in $N$-phonon bundle emission, as shown by analysis of quantum trajectories. Furthermore, we show that $N$-phonon bundle correlation functions display quantum statistics corresponding to anti-bunching for the studied parameters, which is a desired property to realise $N$-phonon guns. Moreover, through the analysis of the emission spectrum, we characterise the acoustic transitions of the system, finding out that $N$-phonon bundle emission can be frequency resolved, which is an important feature to experimentally distinguish the phonon quasiparticles that escape the acoustic cavity.

Usually, high magnetic fields are needed to control the excitation of dark excitons, and producing these fields requires costly and large apparatuses that endanger the scalability possibilities of these systems. The system theoretically studied in this work provides an important alternative for quantum control in acoustic nanocavities without external magnetic fields to emit multi-phonon states through dark exciton manipulation.



\appendix

\section{Experimental Feasibility}\label{sec:feasibility}

So far, we have chosen Hamiltonian and dissipative parameters similar in magnitude to the ones used by~\citet{bin2020n} because they allow us to make direct comparisons between the phenomenology found in their study and ours.
However, these are parameters that are not backed up with the current development of technology, and are distant from an actual experimental realisation of the proposed $N$-phonon bundle emission through the excitation of dark exciton states. 
As an example, it is true that \si{\tera\hertz} acoustic cavities have been built, but they have not been coupled to QDs. 
Instead, experiments such as the one by~\citet{wigger2021} or~\citet{nysten2020} show that the state-of-the-art has only coupled \si{\giga\hertz} acoustic cavities to semiconductor QDs.
In this section, we will show that the phenomenology that we observe still holds for more realistic parameters close to present experiments.

As mentioned, \citet{wigger2021} show that a $\omega_b = 2.9\si{\micro\electronvolt}$ surface acoustic wave (SAW) was coupled to 1.36\si{\electronvolt} In(Ga)As QDs (this is the bare bright exciton energy $\omega_X$) embedded in Bragg mirrors and illuminated by an external laser whose energy can be finely tuned piezoelectrically.
This is exactly the experimental setup needed for the proposal we make in this work.

The internal structure of the QD remains unchanged, meaning that the $\delta$ parameters of the QD (cf.~\cref{eq:qdhamiltonian}) are $\delta_0 = 69.0\omega_b, \delta_1=62.1\omega_b$ and $\delta_2=17.2\omega_b$, which, again, match the values reported by~\citet{bayer2002}.

The quality factor of the SAW determines the phonon escape rate from the phonon cavity defined by the SAW. 
This rate is associated to an energy $\kappa = 10.9\si{\nano\electronvolt}$~\citep{xu2018high}, which can be achieved by further confining the phonon-mode using lateral phononic crystals.

Regarding the external laser, its power can be used to tune the effective laser amplitude $\Omega_1$.
It is well-known that such an amplitude is related to the square of the laser power~\citep{kamada2001,khitrova2006}.
Thus, we set $\Omega_1\approx4.69\omega_b\in[0.1\si{\micro\electronvolt}, 200\si{\micro\electronvolt}]$, where the energy range is a reasonable experimental range according to~\citet{kamada2001}. 
All parameters given in an interval are tuned to minimise the generalised second-order two-phonon bundle correlation function, as explained in~\cite{Note2}.

The dissipative processes involving the QD are exciton decay and dephasing. 
As mentioned in~\cref{sec:results}, the decay rate of dark excitons is lower than that of bright excitons because it involves a random spin flip before spontaneous emission can occur.
Moreover, it has been demonstrated that the exciton decay rates (both bright and dark) can be inhibited or enhanced through optical and electrical control: i.e. embedding the QDs in different optical cavities, and by controlling electric fields with external gates~\citep{dalgarno2005,laucht2009electrical,li2017,lodahl2004}.
Thus, for the bright exciton decay rate, we consider it to be $\gamma_b\approx0.01\omega_b\in [0.01\omega_b, 1.5\omega_b]$. 
This is two orders of magnitude below the assumed value in the experiment by~\citet{wigger2021}. 
Nonetheless, we argue that a better optical cavity that minimises the coupling of the QD bright excitons to optical leaky modes~\citep{perea2004dynamics} (thus, generating an optical band gap) can allow experimental physicists to achieve significantly lower exciton decay rates, such as in the work by~\citet{lodahl2004}. 
As shown in the experiment by~\citet{dalgarno2005}, the dark exciton decay can be tuned to be one or two orders of magnitude lower than that of the bright exciton, which is why we stick to the assumed value in~\cref{sec:results}, i.e. $\gamma_d = 0.1\gamma_b$.
Finally, the pure dephasing rate is assumed to be $\gamma_\phi \approx 0.123\omega_b\in [0.1\omega_b, 2\omega_b]$. 
Pure dephasing is associated to the coupling of the QDs' exciton to the QDs' own lattice phonons~\citep{takagahara1999,besombes2001,muljarov2004}.
The low-temperature broadening has been measured to be in the order of {0.1--1}\si{\micro\electronvolt}~\citep{borri2001,kammerer2002}.
Perhaps, pure dephasing is the parameter that is most difficult to control; however, we point out that recent experiments have shown interesting control mechanisms for pure dephasing in QDs~\citep{liu2015}.

Lastly, the coupling between the confined phonon mode and the QD excitons was assumed to be $g=g_\text{bb}=g_\text{bd} \approx 0.02\omega_b\in[0.01\omega_b, \omega_b]$, which is in the order of tens of nanoelectronvolts. 
Such magnitude has been determined in experiments where phonon-modes are coherently coupled to QDs~\citep{kettler2021}. 
However, we stress that QD coupling to SAW modes can reach larger values, depending on where the QDs are placed: placing them at the nodes of the phonon cavity will yield no coupling, whereas placing them at the anti-nodes will yield the maximum coupling~\citep{Nysten2017,nysten2020}.

\begin{figure}
    \centering
    \includegraphics[width=\columnwidth]{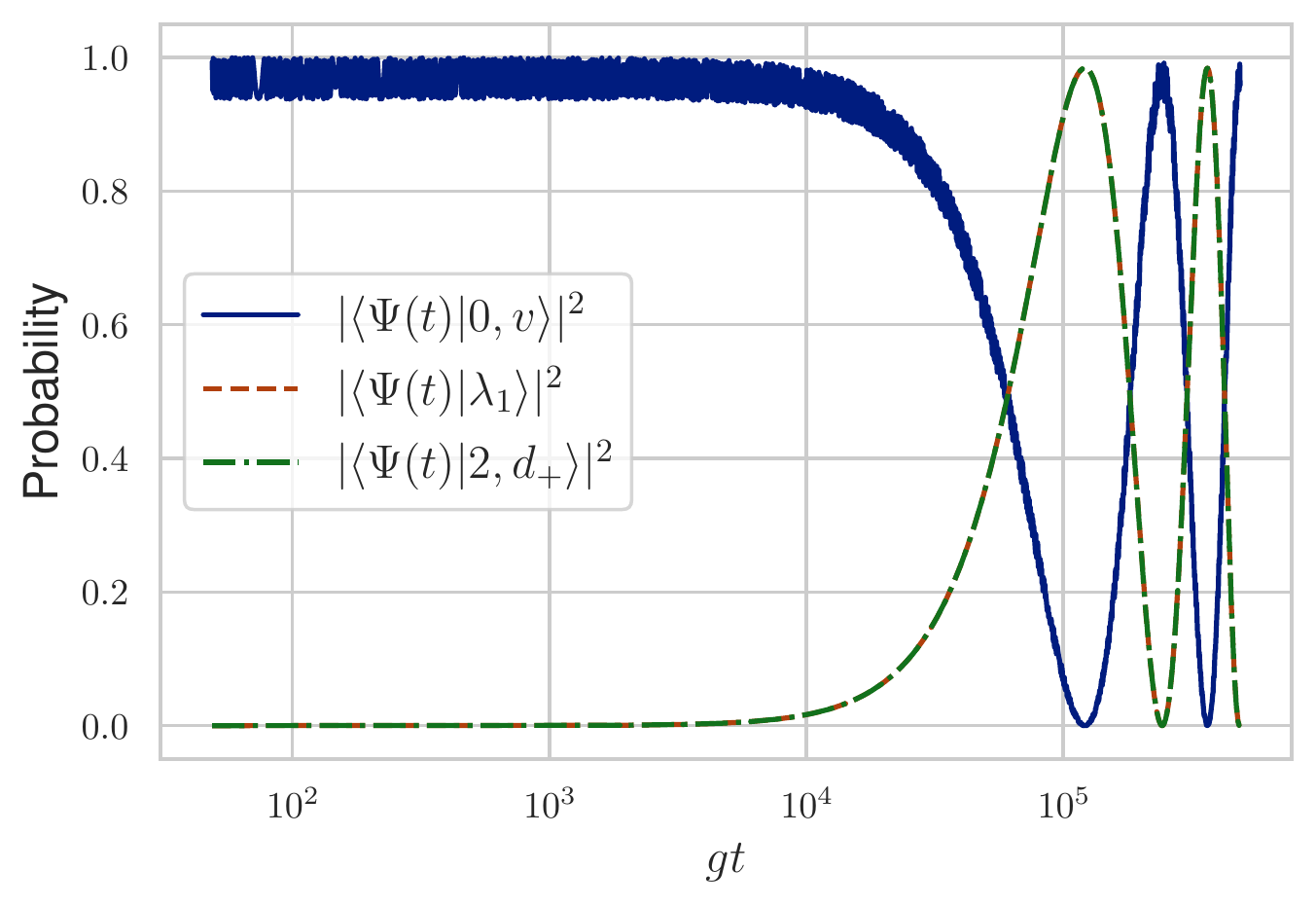}
    \caption{Giant Rabi oscillations between a state $\ket{0,v}$ and $\ket{\lambda_1}\approx\ket{2,d_+}$. As in~\cref{fig:superRabi}, we plot the occupation of the states $\ket{0,v}$ (solid line), $\ket{\lambda_1}$ (dashed line) and $\ket{2,d_+}$ (dot-dashed line). In this regime of strong driving and weak coupling, the laser detuning is $\Delta/\omega_b = (\omega_X - \omega_L)/\omega_b \approx 57.82$.}
    \label{fig:smallfreqRabiOscillations}
\end{figure}

Given the values of the parameters that specify the physical system, we show in~\cref{fig:smallfreqRabiOscillations} that it is also possible to produce giant Rabi oscillations between a state mostly composed of the vacuum state $\ket{0,v}$, and the two-phonon dark symmetric state $\ket{\lambda_1}\approx\ket{2, d_+}$. 
The reason why the $\ket{0,v}$ population shows an additional oscillation is because, for the specified set of parameters, $\ket{0,v}$ is not an eigenstate of the Hamiltonian.
Instead, the giant Rabi oscillations occur, at a specified laser detuning, between the degenerate eigenstates $\ket{\lambda_0}$--which has a dominant component $\ket{0,v}$--and $\ket{\lambda_1}$.
Furthermore, the generalised second-order two-phonon bundle correlation function has a value of $0.39$, which is well within the anti-bunching quantum regime.

\bibliography{apssamp}

\end{document}